\documentclass{iau} 
\usepackage{graphicx}

\title[Mass loss in HB] 
{Mass loss of different
stellar populations in globular clusters}

\author[Marco Tailo]   
{Marco Tailo$^1$}

\affiliation{$^1$Dipartimento di Fisica e Astronomia ``Galileo Galilei'', Univ. di Padova, Vicolo dell'Osservatorio 3, Padova, IT-35122 \\ email: {\tt mrctailo@gmail.com; marco.tailo@unipd.it}}

\pubyear{2019}
\volume{351}  
\jname{Star Clusters: From the Milky Way to the Early Universe}
\editors{A. Bragaglia, M.B. Davies, A. Sills \& E. Vesperini, eds.}
\begin{document}

\maketitle

\begin{abstract}
Once fixed the age and metallicity; the colour distribution of horizontal branch stars in a globular cluster depends on few parameters: the helium abundance of the population and the mass lost during the pre-HB stages. These parameters are usually derived from the HB itself, hence they are degenerate. Breaking this degeneracy and understanding their role is a tricky and challenging problem that no study has tried to solve yet.
Combining the information obtained from the chromosome maps and the analysis of multi-band photometry
with state of the art stellar evolution models, we can obtain a solid estimate of Y for the various stellar populations in a GC. 
We will then have, for the first time, the possibility to break the parameters degeneracy on the HB, understand the role of the mass loss, and lay the foundation to build another piece of the multiple populations mosaic.
\end{abstract}

\firstsection 
\section{Introduction}
The position of a star along the Horizontal Branch (HB) of single age, mono-metallic Globular Clusters (GCs) depends on two parameters: helium abundance and the mass lost during the pre-HB evolutionary stages.  

The evidence that nearly all Galactic GCs host multiple stellar populations with different helium abundance is a major challenge in this context. Indeed, helium-rich stars evolve faster than stars with primordial helium content ($\rm Y \sim 0.25$, Iben  \&  Renzini, 1984). As a consequence, they will produce hotter, less-massive HB stars, which exhibit bluer colours than their lower helium counterparts (e.g.D'Antona et al. 2002). In a similar way, stars that have lost more mass in the pre-HB evolutionary phases will populate the bluer side of the HB. In most papers on the HB modelling, both helium and mass loss are inferred simultaneously from the HB, hence they are degenerate quantities. Small efforts have been done in the past to change this traditional modus operandi and disentangling the role of these two parameters in a GC is a tricky problem that still lack a definitive solution.

Recent work, based on \textit{Hubble Space Telescope} (\textit{HST}) photometry, provided estimates on the helium content of multiple stellar populations in a large sample of GCs (Milone et al. 2018), providing a solid starting point if we want to break the degeneracy between helium and mass loss along the HB.  We apply the method by Milone and collaborators to constrain the helium abundance in M4 and M3 by using their MS and RGB.

\section{The case of M4}

\begin{figure*}
\centering
\includegraphics[width=7cm,height=7cm]{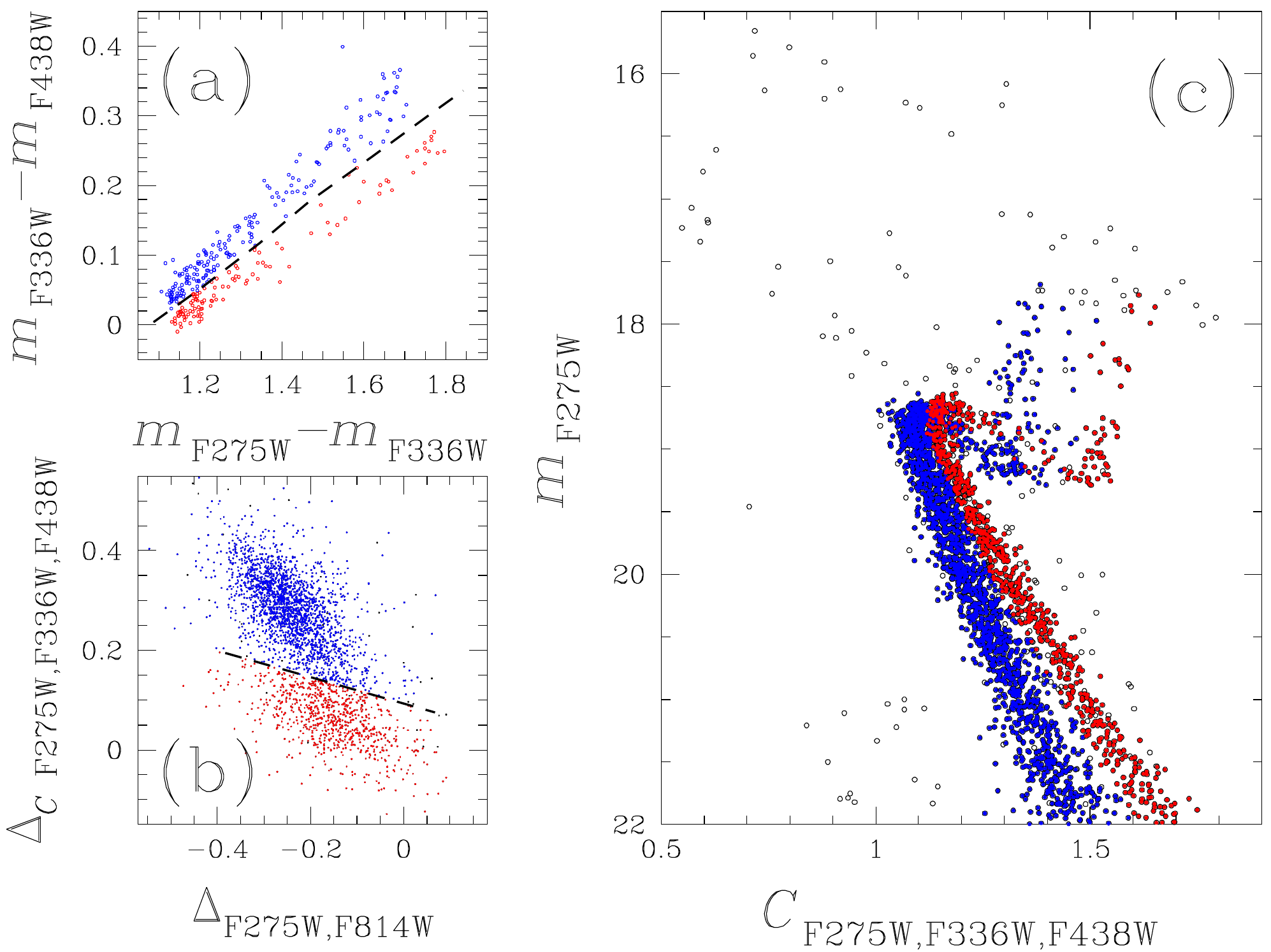}
\includegraphics[width=6cm,height=6.8cm]{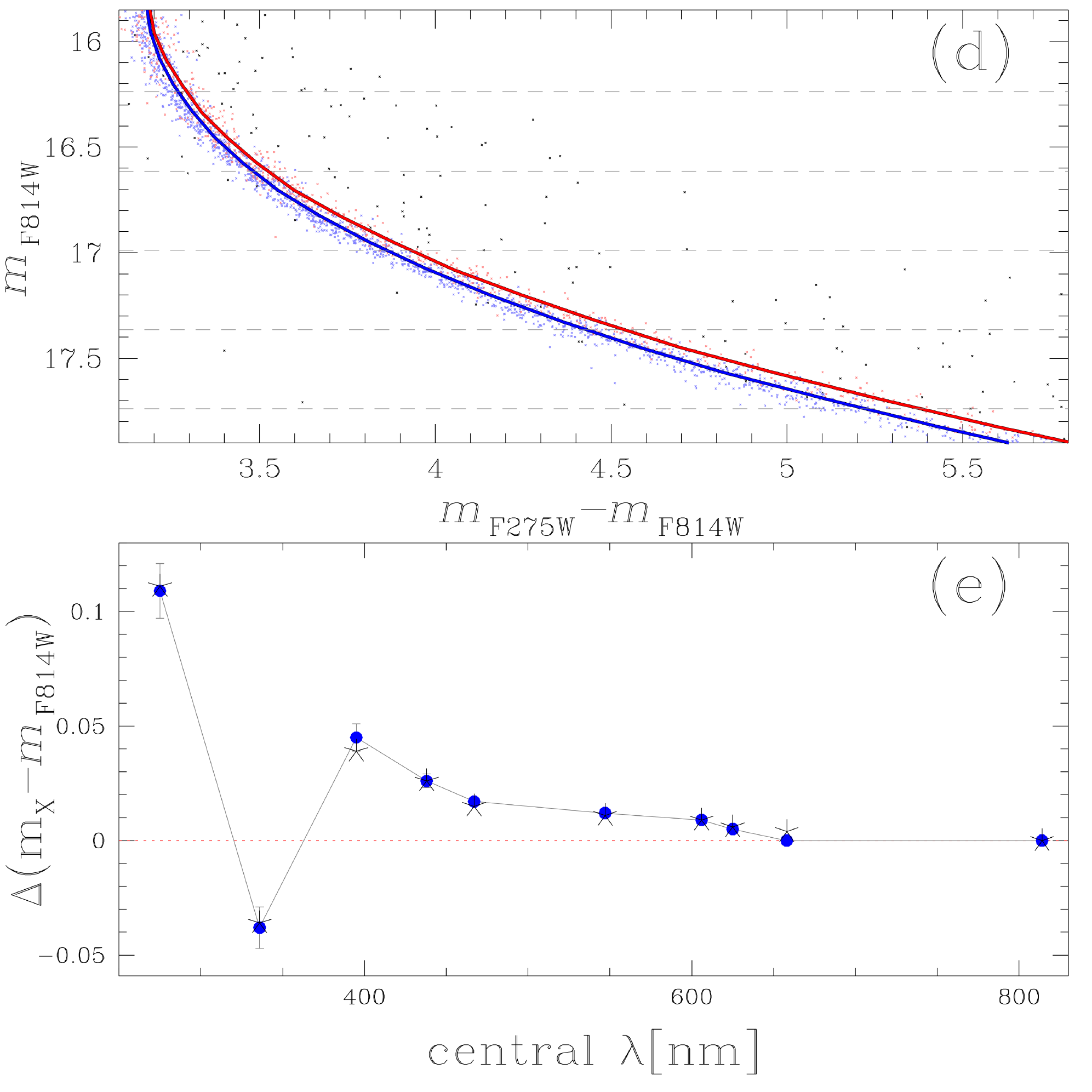}
\caption{A summary of the method to estimate of $\rm \Delta Y$. We separate the two stellar population in M4 using the $m_{\rm F275W}-m_{\rm F336W}$ vs $m_{\rm F336W}-m_{\rm F438W}$ diagram (a), the chromosome map (b) and the $m_{\rm F275W}$ vs $C_{F275W,F336W,F438W}$ diagram (c). We then calculate the colour difference between the fiducials of the two population in multiple bands (d,e). Finally, we translate this colour difference in $\rm \Delta Y$ by using appropriate stellar evolution models and synthetic spectra.}
\label{PIC_M4_1}
\end{figure*}

\begin{figure*}
\centering
\includegraphics[width=11cm,height=7.8cm]{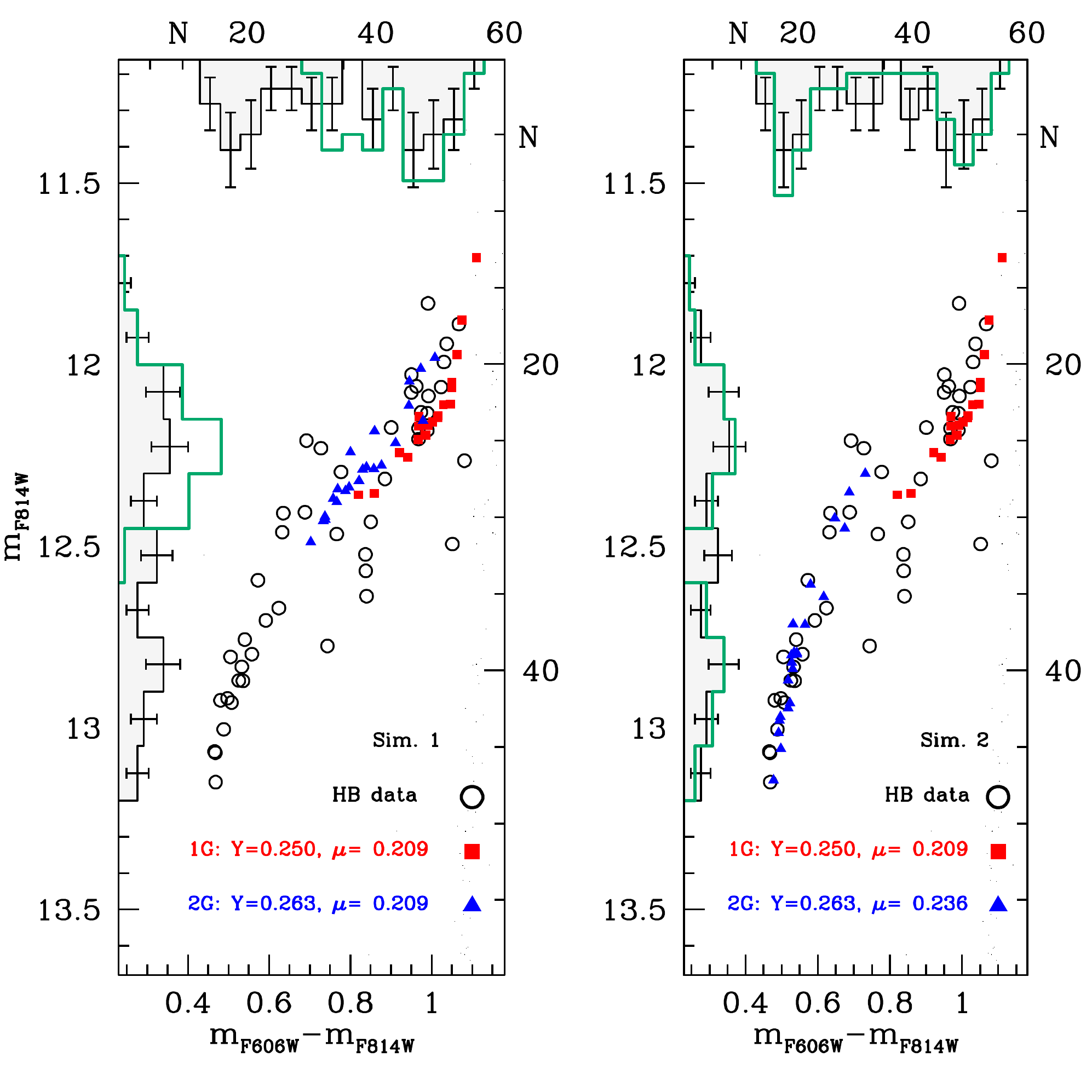}
\caption{\textit{Left panel:} The comparison ot the \textit{HST} data with a simulation obtained using the labelled parameter for Y and $\mu$. The histograms, shaded black for the data and green for the simulation, clearly show that the model does not fit the data well. \textit{Right panel:} The same comparison for a simulation where the 2G stars lose more mass than the 1G ones. In this case we obtain a strong agreement between observed and simulated stars.}
\label{PIC_M4_2}
\end{figure*}

M\,4 is one of the most-studied GC and is the ideal target for our purpose due to its simplicity.  High-resolution spectroscopy revealed two main stellar populations: a first-generation (1G) with low sodium ([Na/Fe] $\sim$ 0.1) and high oxygen ([O/Fe]$\sim$0.5) and a second stellar generation (2G) enhanced in sodium ([Na/Fe]$\sim$0.45) with lower oxygen ([O/Fe]$\sim$0.2) (see Marino et al. 2008). These two populations can be identified in the entire colour-magnitude diagram (CMD), from the RGB (Marino et al. 2008) to the bottom of the main sequence (Milone et al. 2014). The HB of NGC\,6121 is also bi-modal and is well populated on both sides of the instability strip. Moreover, high-resolution spectroscopy (Marino et al. 2011) shows that red and blue HB stars belong to the 1G and 2G, respectively, providing an additional constraint.

We applied to M4 the procedure introduced by Milone et al. (2012) and Milone et al. (2018). The procedure is summarized in Fig. \ref{PIC_M4_1}.
Once the populations have been separated (see Fig. \ref{PIC_M4_1}a, \ref{PIC_M4_1}b, \ref{PIC_M4_1}c), we derive the MS fiducial lines of 1G and 2G stars in each CMD (e.g like the one in Fig. \ref{PIC_M4_1}d). We then calculate the colour differences between the fiducials of 2G and 1G stars for all the available filters (see Fig. \ref{PIC_M4_1}e). We combine these results with a grid of synthetic spectra and stellar evolution models to infer the value of helium enhancement (see Milone et al. 2018). We obtain for M4 $\rm \Delta Y = 0.013 \pm 0.002$.

With this new constraint, we then simulate the HB of M4. Our results are summarized in Fig. \ref{PIC_M4_2}. We know from spectroscopy that the 1G stars populate the red HB. We thus obtain an estimate of the total mass loss ($\mu$) necessary to reproduce their position. We find $\rm \mu_{1G}= 0.209\, M/M_\odot$. Note that we approximate the primordial helium abundance to Y=0.25. If we use the same value to simulate the 2G stars (Simulation 1 shown in the left panel of Fig \ref{PIC_M4_2}) we obtain a bad fit, as the the comparison of the histograms in the figure shows. On the other hand, increasing the value of $\mu$ allows for a better reproduction of the observed colour distribution, as reported in the right panel of Fig. \ref{PIC_M4_2}. The difference in $\mu$ between the two generations of stars needed is $\rm \Delta \mu_{2G,1G} = 0.027 \pm 0.006\, M/M_\odot$.

\section{The case of M3}

The GC M3 hosts a handful of stellar populations, and multiple groups of stars can be identified, from the chromosome map, in its 2G  (Milone et al. 2017) . Moreover the large number of RRLyrae variables (Benk{\H{o}} et al. 2006) hosted in this cluster offers an additional strong constraint on the HB colour distribution. Given the complexity of its patterns, we divide the 2G stars in two main groups: one ($\rm 2G_A$) containing the bulk of the 2G stars and a second ($\rm 2G_B$) located the farthest form the 1G on the map.  

\begin{figure*}
\centering
\includegraphics[width=11cm,height=7.8cm]{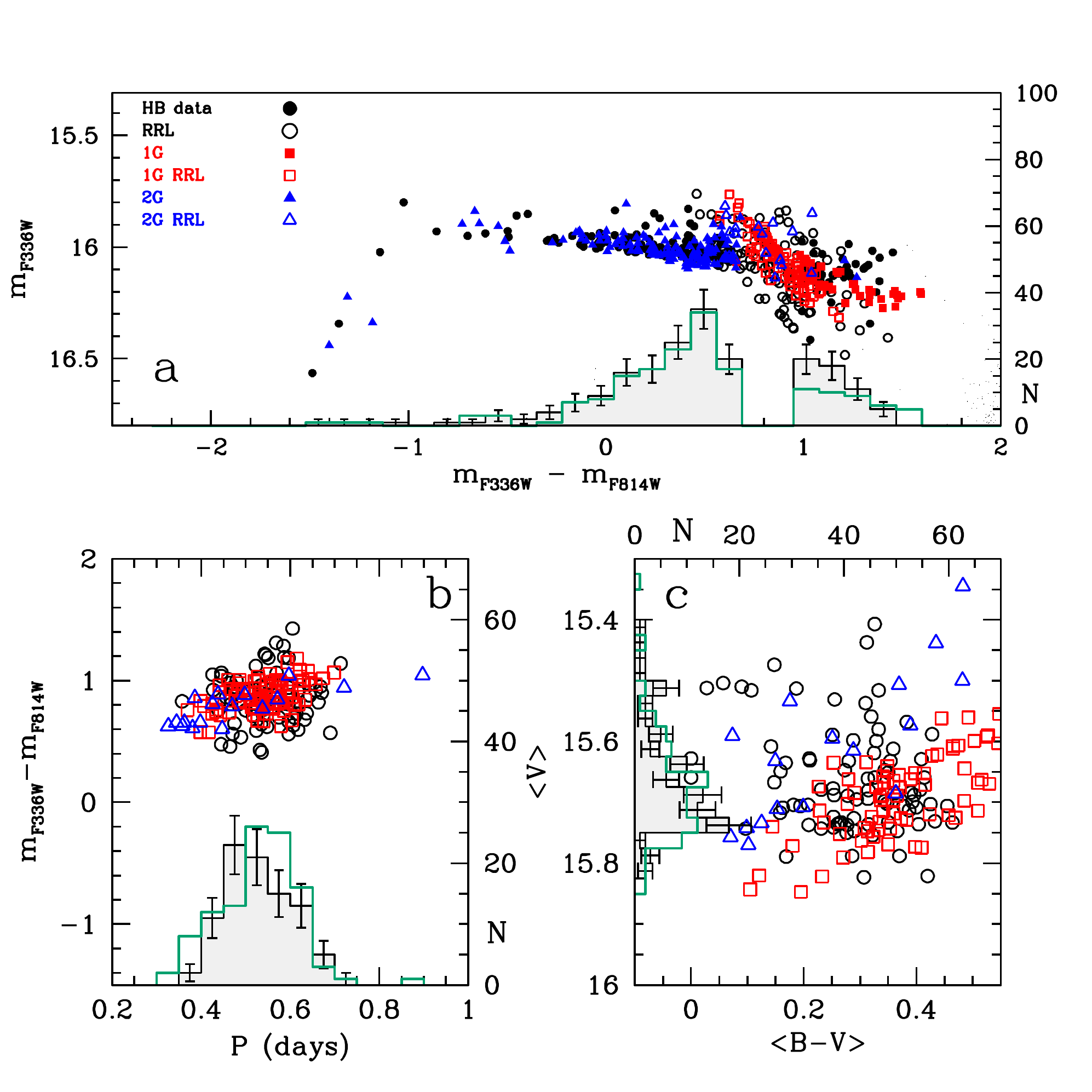}
\caption{\textit{Panel a:} Comparison of the \textit{HST} data with a simulation where the 2G stars lose more mass than the 1G. The comparison between the histograms, color coded as in Fig. \ref{PIC_M4_2} suggest a strong agreement between the data and the simulation. \textit{Panel b an c:} the comparison of the period and mean magnitude distribution of the RRLyrae stars in the \textit{HST} data with the ones in the simulated HB. We reach a satisfactory agreement with the data.}
\label{PIC_M3}
\end{figure*}

We use then the results by Milone et al. (2018) for M3 ($\rm \Delta Y^{2G,1G} = 0.016$ and $\rm \Delta Y_{max}^{2G,1G} = 0.041$) and model the entire HB of this GC (Fig. \ref{PIC_M3}a). We locate the 1G stars on the reddest part of the HB, obtaining for them $\mu_{1G}=0.188\,M/M_\odot$.  Since Y is measured independently from the HB morphology, we are able to obtain, also in this complex case, a measure of the mass lost of the 2G stars. In order to get a good fit of the entire HB extension we need to increase the mass loss by $\rm \Delta \mu^{2GA,1G} = 0.016\pm 0.007\,  M/M_\odot$ and $\rm \Delta \mu^{2GB,1G} = 0.052 \pm 0.013\, M/M_\odot$ between the 1G and the groups that form the 2G. The simulation also achieve the description of the period and magnitude distribution of the large number of RR Lyrae variables found in this cluster (panels b and c).

\section{Conclusions}

We introduced a new method to infer the mass loss of 1G and 2G stars, and disentangle the role of helium and mass loss on the HB morphology. Once we have estimated the relative helium abundance of multiple populations from multi-band photometry of MS or RGB stars, we can measure the mass loss of 1G and 2G stars needed to reproduce the entire HB.

We applied this method to M3 and M4 and find that in both clusters we
need to increase the mass loss of the 2G to be able to describe the blue-most stars in the HB. The amount of additional mass the 2G stars need to lose is too high to be connected to the structural differences alone, given the small helium abundance differences involved. Other, still unknown, mechanisms are needed in order to achieve such additional mass loss. We refer the interested reader
to the papers by Tailo et al. (2019a, 2019b) for further details.

\section*{Acknowledgments}
MT acknowledges support the European Research Council (ERC) under the European Union's Horizon 2020 research innovation programme (Grant Agreement ERC-StG 2016, No 716082 'GALFOR', PI: Milone) and from MIUR through the the FARE project R164RM93XW ‘SEMPLICE’ (PI: Milone) and  .

\end{document}